\documentclass[aps,prb,reprint,superscriptaddress,longbibliography]{revtex4-2}
\usepackage[utf8]{inputenc}
\usepackage{graphicx}
\usepackage{amsmath,amssymb,bm}
\begin{document}

\title{Hydrostatic Pressure-Induced Evolution of the Superconducting Transition Temperature of Bi-2212: Insights from First-Principles Calculations}

\author{Shuhong Tang}
\affiliation{Science Island Branch of Graduate School, University of Science and Technology of China, Hefei 230026, China}
\affiliation{Key Laboratory of Materials Physics, Institute of Solid State Physics, HFIPS, Chinese Academy of Sciences, Hefei 230031, China}
\author{Hanyu Wang}
\affiliation{Science Island Branch of Graduate School, University of Science and Technology of China, Hefei 230026, China}
\affiliation{Key Laboratory of Materials Physics, Institute of Solid State Physics, HFIPS, Chinese Academy of Sciences, Hefei 230031, China}
\author{Di Peng}
\affiliation{Science Island Branch of Graduate School, University of Science and Technology of China, Hefei 230026, China}
\affiliation{Key Laboratory of Materials Physics, Institute of Solid State Physics, HFIPS, Chinese Academy of Sciences, Hefei 230031, China}
\author{Da-yong Liu}
\affiliation{Department of Physics, School of Physical Science and Technology, Nantong University, Nantong 226019, China}
\author{Zhi Zeng}
\thanks{Corresponding author.}
\affiliation{Science Island Branch of Graduate School, University of Science and Technology of China, Hefei 230026, China}
\affiliation{Key Laboratory of Materials Physics, Institute of Solid State Physics, HFIPS, Chinese Academy of Sciences, Hefei 230031, China}
\author{Liang-Jian Zou}
\thanks{Corresponding author.}
\affiliation{Science Island Branch of Graduate School, University of Science and Technology of China, Hefei 230026, China}
\affiliation{Key Laboratory of Materials Physics, Institute of Solid State Physics, HFIPS, Chinese Academy of Sciences, Hefei 230031, China}

\begin{abstract}
High-pressure experiments on Bi$_2$Sr$_2$CaCu$_2$O$_{8+x}$ (Bi-2212) have reported apparently conflicting evolutions of the superconducting transition temperature $T_{c}$, ranging from weak enhancement to strong suppression and even a proposed second superconducting dome. To clarify the origin of these discrepancies, we combine first-principles density functional theory (DFT) calculations with a pressure-dependent low-energy bilayer model solved by the slave-boson mean-field (SBMF) method together with a Berezinskii-Kosterlitz-Thouless estimate of phase coherence. Our results show that hydrostatic pressure induces a pronounced self-doping effect in Bi-2212: holes are transferred from the Bi-O charge-reservoir layers to the CuO$_2$ superconducting planes, leading to a systematic increase in the effective CuO$_2$-plane hole concentration $\delta_x$. At the same time, pressure enhances the pairing scale through the renormalization of the hopping and superexchange parameters. As a consequence, the pressure evolution of $T_{c}$ is governed by the competition between pressure-enhanced pairing and pressure-driven motion along the common $T_{c}$-$\delta_x$ dome, making $T_{c}(P)$ extraordinarily sensitive to the initial doping state. Even samples with very similar ambient-pressure $T_{c}$ but slightly different initial doping can therefore display qualitatively different pressure responses. This provides a unified interpretation of a large part of the disparate high-pressure behavior reported for Bi-2212 and suggests that slightly underdoped samples are more favorable than ambient-pressure optimal samples for achieving improved superconducting performance under pressure.
\end{abstract}

\maketitle

\section{Introduction}

Among the various experimental tuning parameters, hydrostatic pressure has long been a particularly effective route to enhance the superconducting transition temperature ($T_{c}$) \cite{ref1,ref2}. Yet the pressure response can be highly nontrivial, especially in Bi-based cuprates \cite{ref3,ref4,ref5,ref6,ref7,ref8}. The Bi-family cuprate superconductors with the general formula Bi$_2$Sr$_2$Ca$_{n-1}$Cu$_n$O$_{2n+4+x}$ (Bi-2201, Bi-2212, and Bi-2223 for $n = 1$, 2, 3, where $x$ denotes the oxygen content) have been extensively studied for both their high-$T_{c}$ phenomenology and potential applications \cite{ref9,ref10,ref11}. Despite substantial experimental efforts under pressure \cite{ref3,ref4,ref5,ref6,ref7,ref8}, the experimental evolution of $T_{c}$ remains unsettled. Adachi \textit{et al.} \cite{ref8} found a single dome in Bi-2212, with $T_{c}$ initially increasing and then being suppressed at higher pressures toward a normal metallic state. By contrast, Deng \textit{et al.} \cite{ref4} reported an unusual two-dome evolution of $T_{c}$ under pressure in both Bi-2212 and Bi-2223, a behavior not established in other cuprates. More recently, Zhou \textit{et al.} \cite{ref5} suggested that, following the high-pressure suppression of superconductivity, Bi-2212 enters an insulating-like regime, indicative of a possible superconductor--insulator transition. These discrepancies call for a systematic theoretical framework to disentangle the intertwined roles of pressure and doping in Bi-based cuprates.

To address this issue, we model the pressure- and doping-dependent electronic structure of Bi-2212 using first-principles density functional theory (DFT) calculations, and track the evolution of the band structure over a broad parameter range. Given the strong correlations associated with Cu $3d$ electrons, we further analyze the superconducting instability within a slave-boson mean-field treatment supplemented by a Berezinskii--Kosterlitz--Thouless (BKT) phase-ordering criterion, enabling a quantitative determination of $T_{c}$ as a function of pressure and doping.

Our main result is that hydrostatic pressure induces a pronounced self-doping effect in Bi-2212 and makes the pressure evolution of $T_{c}$ extraordinarily sensitive to the initial doping state. Pressure increases the effective hole concentration in the CuO$_2$ planes, but whether this enhances or suppresses $T_{c}$ depends on where the sample starts on the common $T_{c}$-$\delta_x$ dome. Consequently, even near-optimal samples with very similar ambient-pressure $T_{c}$ can evolve very differently under compression. This provides a natural way to reconcile a large part of the apparent disagreement among high-pressure experiments and suggests that slightly underdoped samples are more favorable than ambient-pressure optimal samples for achieving improved superconducting performance under pressure. Section II describes the computational methodology. Section III presents the structural, electronic, and superconducting results under pressure. Section IV discusses their implications and summarizes the main conclusions.

\section{Theoretical Calculation Methods}

In this section we briefly summarize the methods used to determine the pressure-dependent crystal structure and electronic structure and to estimate $T_{c}$.

\subsection{First-Principles Method for Electronic Structure}

All structural optimizations under hydrostatic pressure, including the determination of equilibrium lattice parameters, were performed within density functional theory (DFT) using the Vienna \textit{Ab initio} Simulation Package (VASP) \cite{ref12}. The exchange--correlation potential was treated within the generalized gradient approximation in the Perdew--Burke--Ernzerhof (PBE) form \cite{ref13}, and ion--electron interactions were described by the projector augmented-wave (PAW) method \cite{ref14}. A plane-wave cutoff energy of 600 eV was used throughout. Structural relaxations were carried out using a $\Gamma$-centered $7\times 7\times 3$ $k$-point mesh. All atomic positions were fully relaxed until the total-energy change was below $10^{-8}$ eV and the residual Hellmann--Feynman forces were less than 0.001 eV/\AA. The optimized structures at each pressure were then taken as input for all-electron electronic-structure calculations using the WIEN2k code \cite{ref15}, which implements the full-potential linearized augmented plane-wave plus local orbitals (FP-LAPW+lo) method and is widely regarded as a benchmark for high-accuracy band-structure calculations. The following atomic sphere radii were used consistently across all pressures: 2.0, 2.2, 2.1, 1.7, and 1.5 a.u. for Bi, Sr, Ca, Cu, and O, respectively. The core--valence separation was set by a cutoff energy of $-11$ Ry. The exchange--correlation potential was treated within the generalized gradient approximation using the Perdew--Burke--Ernzerhof (PBE) functional \cite{ref13}. Brillouin-zone sampling employed a $10\times 10\times 10$ $k$-point grid for self-consistent calculations and a denser $15\times 15\times 15$ grid for density-of-states and electronic-structure analyses. Self-consistency was achieved with a total-energy convergence criterion of $10^{-5}$ eV.

In the CuO$_2$ planes, the low-energy electronic degrees of freedom are primarily derived from the Cu $d_{x^{2}-y^{2}}$ and O $p_{x}/p_{y}$ orbitals, which hybridize to form the $pd$-$\sigma$ manifold. This hybridization yields a characteristic band crossing the Fermi level that is widely viewed as central to cuprate superconductivity \cite{ref16,ref17,ref18,ref19}. In the undoped limit, this band is nominally half-filled (occupancy $\sim 1$ per Cu). Carrier doping shifts the chemical potential and changes the band filling, in qualitative correspondence with the experimentally observed emergence of superconductivity. Accordingly, variations in the occupancy of the relevant low-energy band(s) can be interpreted as changes in the hole concentration.

In our simulations, the crystal structures, including both lattice parameters and atomic positions, were fully relaxed at each pressure. This approach allows us to isolate the electronic response to purely hydrostatic pressure, without contributions from non-hydrostatic stresses that may occur in experiments. To model doping effects, we adopted the rigid-band approximation, wherein doping is assumed to shift the Fermi level without significantly altering the underlying band structure. Within this approximation, the mapping from nominal oxygen doping to the effective CuO$_2$-plane hole concentration should be regarded as approximate.

\subsection{The Low-Energy Effective Model for Bi-2212}

In Bi-2212, the electronic structure near the Fermi level is dominated by the Cu $d_{x^{2}-y^{2}}$ orbitals within the bilayer CuO$_2$ planes. The corresponding low-energy dispersion can therefore be captured by a bilayer single-orbital model \cite{ref20}. The overlap between neighboring $d_{x^{2}-y^{2}}$ orbitals gives rise to significant intra- and interlayer hopping, described by the tight-binding Hamiltonian

\begin{equation}
\begin{split}
H_{0}={}&-\sum_{ij,\alpha,\sigma}'
t_{\parallel,ij}\left(d_{\alpha i\sigma}^{\dagger}d_{\alpha j\sigma}+\mathrm{h.c.}\right) \\
&-t_{\perp}\sum_{i,\sigma}\left(d_{1i\sigma}^{\dagger}d_{2i\sigma}+\mathrm{h.c.}\right),
\end{split}
\end{equation}

where $i$ labels the lattice site, $\alpha$ the layer index, and $\sigma$ the spin. Here $d_{\alpha i\sigma}^{\dagger}$ creates an electron in the Cu $d_{x^{2}-y^{2}}$ orbital, and $\sum_{ij}'$ denotes a restricted sum over first- and second-nearest-neighbor intralayer pairs.

As a minimal description of the local correlation on this projected Cu $d_{x^{2}-y^{2}}$ manifold, we consider the onsite Hubbard interaction

\begin{equation}
H_{U}=U\sum_{\alpha i}n_{\alpha i\uparrow}n_{\alpha i\downarrow}.
\end{equation}

In the strong-coupling limit, the projected low-energy Hamiltonian reduces to an effective bilayer single-orbital $t$-$J$ model, which can be treated within the slave-boson mean-field framework to study superconductivity \cite{ref21}:

\begin{equation}
H=H_{0}+H_{\mathrm{int}}
\end{equation}

\begin{equation}
H_{\mathrm{int}}=J\sum_{\langle i,j\rangle,\alpha}\mathbf{S}_{\alpha i}\cdot\mathbf{S}_{\alpha j}.
\end{equation}

In this model we retain only the intralayer superexchange interaction $J$, neglecting the interlayer term because of its relatively small magnitude. Consistent with the strong-correlation limit, double occupancy is prohibited, so the interaction is represented solely by the $J$ term in Eq.~(4). Similar effective models have been widely employed in studies of cuprates and are well validated \cite{ref20,ref21,ref22,ref23}.

For each pressure, the hopping and exchange parameters of the model must be determined individually. The hopping parameters are obtained primarily through Wannier-function fitting of the first-principles electronic structures using the Wannier90 package \cite{ref24}. This procedure ensures that the effective low-energy model accurately reproduces the \textit{ab initio} band structure and provides a reliable basis for introducing the interaction terms. The superexchange parameter $J$ is then estimated from the kinetic scale through the standard relation $J\sim 4t^{2}/U$ \cite{ref21}. In the present work, we keep $U$ fixed and therefore assume that the dominant pressure dependence of $J$ follows $t^{2}/U$, i.e., it is controlled primarily by the evolution of the hopping amplitude, while other possible pressure-induced contributions to the superexchange are neglected.

\subsection{The SBMF Method and the Estimation of $T_{c}$}

To enforce the no-double-occupancy constraint, we employ the Barnes--Coleman slave-boson method \cite{ref21,ref25,ref26} on the effective low-energy model in Eq.~(3). Within the slave-boson mean-field (SBMF) framework, the electron creation operator $d_{\alpha i\sigma}^{\dagger}$ is decomposed into a bosonic operator $b_{\alpha i}$ (holon annihilation operator) and a fermionic operator $f_{\alpha i\sigma}^{\dagger}$ (spinon creation operator) via

\begin{equation}
d_{\alpha i\sigma}^{\dagger}=b_{\alpha i}f_{\alpha i\sigma}^{\dagger}
\end{equation}

The constraint against double occupancy is enforced by introducing an auxiliary condition, expressed as

\begin{equation}
b_{\alpha i}^{\dagger}b_{\alpha i}+\sum_{\sigma}f_{\alpha i\sigma}^{\dagger}f_{\alpha i\sigma}=1
\end{equation}

which must be satisfied at every site and layer.

This constraint is enforced via a Lagrange multiplier ${\lambda}_{\alpha}$, which acts as the chemical potential of the holon field. The electron occupation number, determined self-consistently from first-principles calculations at various pressures, serves as a key input. In the undoped case, each $d_{x^{2}-y^{2}}$ orbital has an occupation number of one, corresponding to the half-filled Mott insulating state. Under pressure, we find that the occupation number varies between 0.7 and 1.0 per orbital, reflecting changes in the hole doping level within the actual material.

In the SBMF approximation, $H_{0}$ can be expressed as:

\begin{equation}
\begin{split}
H_{0}={}&\sum_{ij,\alpha,\sigma}'
-t_{\parallel,ij}\Bigl(
b_{\alpha j}^{\dagger}b_{\alpha i}f_{\alpha i\sigma}^{\dagger}f_{\alpha j\sigma} \\
&\quad+\mathrm{h.c.}\Bigr) \\
&-t_{\perp}\sum_{i,\sigma}\left(b_{2i}^{\dagger}b_{1i}f_{1i\sigma}^{\dagger}f_{2i\sigma}+\mathrm{h.c.}\right)
\end{split}
\end{equation}

We introduce the spinon and holon bond fields

\begin{equation}
\begin{aligned}
\chi_{\parallel}={}&\sum_{\sigma}\left\langle
f_{\alpha i\sigma}^{\dagger}f_{\alpha,i+\hat{x},\sigma}\right\rangle \\
={}&\sum_{\sigma}\left\langle
f_{\alpha i\sigma}^{\dagger}f_{\alpha,i+\hat{y},\sigma}\right\rangle, \\
\chi_{\parallel}'={}&\sum_{\sigma}\left\langle f_{\alpha i\sigma}^{\dagger}
f_{\alpha,i+\hat{x}+\hat{y},\sigma}\right\rangle, \\
\chi_{\perp}={}&\sum_{\sigma}\left\langle f_{1i\sigma}^{\dagger}f_{2i\sigma}\right\rangle
\end{aligned}
\end{equation}

\begin{equation}
\begin{aligned}
\kappa_{\parallel}={}&\left\langle b_{\alpha i}^{\dagger}b_{\alpha,i+\hat{x}}\right\rangle
=\left\langle b_{\alpha i}^{\dagger}b_{\alpha,i+\hat{y}}\right\rangle, \\
\kappa_{\parallel}'={}&\left\langle b_{\alpha i}^{\dagger}
b_{\alpha,i+\hat{x}+\hat{y}}\right\rangle, \\
\kappa_{\perp}={}&\left\langle b_{1i}^{\dagger}b_{2i}\right\rangle
\end{aligned}
\end{equation}

together with the intralayer $d_{x^{2}-y^{2}}$-wave spinon pairing amplitude

\begin{equation}
\begin{split}
\Delta_{d}=\Delta_{x}={}&\left\langle
f_{\alpha i\uparrow}f_{\alpha,i+\hat{x},\downarrow}
-f_{\alpha i\downarrow}f_{\alpha,i+\hat{x},\uparrow}\right\rangle \\
={}&-\left\langle
f_{\alpha i\uparrow}f_{\alpha,i+\hat{y},\downarrow}
-f_{\alpha i\downarrow}f_{\alpha,i+\hat{y},\uparrow}\right\rangle \\
={}&-\Delta_{y}
\end{split}
\end{equation}

These MF order parameters are assumed to be site/layer independent MF values and are taken to be real.

The spin superexchange interaction term $H_{int}$ between adjacent lattice sites can be decoupled as:

\begin{equation}
\begin{aligned}
J\mathbf{S}_{\alpha i}\cdot\mathbf{S}_{\alpha j}\simeq{}&
\frac{3J}{8}\left(\left|\chi_{\parallel}\right|^{2}+\left|\Delta_{d}\right|^{2}\right) \\
&-\frac{3J}{8}\Bigl[\chi_{\parallel}\left(f_{\alpha i\uparrow}^{\dagger}f_{\alpha j\uparrow}+f_{\alpha i\downarrow}^{\dagger}f_{\alpha j\downarrow}\right)+\mathrm{h.c.}\Bigr] \\
&-\frac{3J}{8}\Bigl[\Delta_{d}\left(f_{\alpha i\uparrow}^{\dagger}f_{\alpha j\downarrow}^{\dagger}-f_{\alpha i\downarrow}^{\dagger}f_{\alpha j\uparrow}^{\dagger}\right)+\mathrm{h.c.}\Bigr]
\end{aligned}
\end{equation}

In the original layer--Nambu basis ${\Phi}_{k}^{\dagger}=\left(f_{1k\uparrow}^{\dagger},f_{2k\uparrow}^{\dagger},f_{1-k\downarrow},f_{2-k\downarrow}\right)$, the spinon Bogoliubov--de Gennes Hamiltonian has a $4\times 4$ form. Because the two layers are equivalent and only the normal interlayer hopping $t_{\perp}$ is retained, a bonding/antibonding transformation diagonalizes the layer sector and reduces the problem to two independent $2\times 2$ blocks labeled by $\eta=\pm1$,

\begin{equation}
\begin{split}
H_{\mathrm{MF}}={}&\sum_{k,\eta}\omega_{k\eta}\,b_{k\eta}^{\dagger}b_{k\eta} \\
&+\sum_{k,\eta}\Psi_{k\eta}^{\dagger}
\begin{pmatrix}\xi_{k\eta} & \Delta_{k} \\ \Delta_{k} & -\xi_{k\eta}\end{pmatrix}
\Psi_{k\eta} \\
&+N\Omega_{0}
\end{split}
\end{equation}

with ${\Psi}_{k\eta}^{\dagger}=\left(f_{k\eta\uparrow}^{\dagger},f_{-k\eta\downarrow}\right)$ and

\begin{equation}
\begin{split}
\omega_{k\eta}={}&\lambda-2t\chi_{\parallel}\left(\cos k_{x}+\cos k_{y}\right) \\
&-4t'\chi_{\parallel}'\cos k_{x}\cos k_{y}-\eta t_{\perp}\chi_{\perp},
\end{split}
\end{equation}

\begin{equation}
\begin{split}
\xi_{k\eta}={}&E_{0}+\lambda-\mu \\
&-2\left(t\kappa_{\parallel}+\frac{3J}{8}\chi_{\parallel}\right)
\left(\cos k_{x}+\cos k_{y}\right) \\
&-4t'\kappa_{\parallel}'\cos k_{x}\cos k_{y}
-\eta t_{\perp}\kappa_{\perp},
\end{split}
\end{equation}

\begin{equation}
\begin{aligned}
\Delta_{k}&=\frac{3J}{4}\Delta_{d}\left(\cos k_{x}-\cos k_{y}\right), \\
E_{k\eta}&=\sqrt{\xi_{k\eta}^{\,2}+\Delta_{k}^{\,2}}
\end{aligned}
\end{equation}

The hole density $\delta_x$ in the CuO$_2$ planes is fixed by

\begin{equation}
\begin{aligned}
2\delta_{x}={}&\sum_{\alpha}\left\langle b_{\alpha i}^{\dagger}b_{\alpha i}\right\rangle, \\
2\left(1-\delta_{x}\right)={}&\sum_{\alpha,\sigma}
\left\langle f_{\alpha i\sigma}^{\dagger}f_{\alpha i\sigma}\right\rangle
\end{aligned}
\end{equation}

At fixed hole concentration $\delta_x$ on the CuO$_2$ planes, the saddle-point parameters ${{\chi}_{\parallel},{\chi}_{\parallel}',{\chi}_{\perp},{\kappa}_{\parallel},{\kappa}_{\parallel}',{\kappa}_{\perp},{\Delta}_{d},\lambda,\mu}$ are determined self-consistently from the order-parameter definitions together with the density constraints. In the numerical implementation, the coupled boson and spinon equations are iterated until full self-consistency is reached. The temperature at which $\Delta_d(T)$ vanishes defines the pair-formation scale $T_{pair}$.

Owing to the two-dimensional character of cuprate superconductors, the superconducting transition temperature is ultimately governed by the Berezinskii--Kosterlitz--Thouless (BKT) superfluid phase transition \cite{ref27}. To estimate both normal and superconducting properties, we follow the approach described in Ref.~\cite{ref28}, which incorporates the slave-boson mean-field pairing order parameter in a manner analogous to BCS theory, while also accounting for the BKT transition. We therefore evaluate the phase stiffness explicitly. The holon contribution is obtained from the curvature of the phase-twisted condensate free energy $F_{b}\left(q\right)$. For a uniform twist $q=(q_{x},q_{y})$, the holon stiffness

\begin{equation}
\begin{split}
K_{b}(T)=\frac{1}{2}\left[
\frac{\partial^{2}F_{b}(q)}{\partial q_{x}^{2}}
+\frac{\partial^{2}F_{b}(q)}{\partial q_{y}^{2}}
\right]_{q=0}
\end{split}
\end{equation}

The spinon pair-phase stiffness is computed in the same manner from the curvature of the twisted Bogoliubov free energy ${\Omega}_{f}(q)$,

\begin{equation}
\begin{split}
K_{f}(T)=\frac{1}{2}\left[
\frac{\partial^{2}\Omega_{f}(q)}{\partial q_{x}^{2}}
+\frac{\partial^{2}\Omega_{f}(q)}{\partial q_{y}^{2}}
\right]_{q=0}
\end{split}
\end{equation}

Following the Ioffe--Larkin composition rule \cite{ref29}, the physical superfluid stiffness is constructed as

\begin{equation}
K_{\mathrm{phys}}(T)=\frac{K_{b}(T)K_{f}(T)}{K_{b}(T)+4K_{f}(T)}
\end{equation}

where the factor $4$ follows from the normalization of the charge-$2$ spinon-pair phase field adopted in the present formulation.

Finally, the characteristic temperatures are determined from the BKT criterion,

\begin{equation}
T_{c}=\frac{\pi}{2}K_{\mathrm{phys}}(T_{c})
\end{equation}

In the following, the superconducting transition temperature quoted for Bi-2212 is the physical BKT scale $T_{c}$, while $T_{pair}$ only marks the onset of spinon pairing. This formulation naturally separates pair formation from phase ordering and is therefore well suited to the underdoped regime, where phase fluctuations can substantially suppress $T_{c}$ relative to the pairing scale. All pressure-dependent input parameters and the corresponding CuO$_2$-plane hole concentrations are taken from the first-principles analysis described above.

\begin{figure*}[t]
\centering
\includegraphics[width=\textwidth]{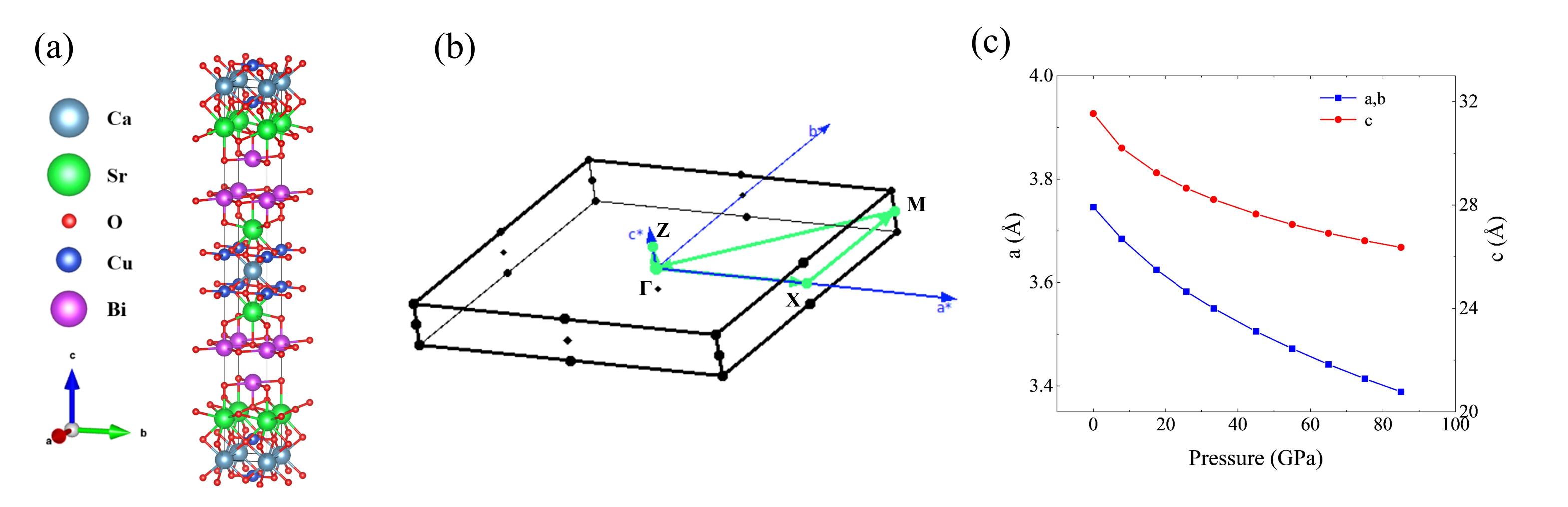}
\caption{(a) Crystal structure of Bi-2212 in the body-centered tetragonal approximation (I4/mmm). (b) Corresponding Brillouin zone and high-symmetry path used in the band-structure calculations. (c) Pressure dependence of the optimized lattice parameters obtained from DFT structural relaxations.}
\label{fig:figure1}
\end{figure*}

\section{Pressure-Induced Evolution of Structure and Electronic States}

A central task in understanding the superconducting properties of Bi-2212 under pressure is to determine how both the crystal structure and the low-energy electronic structure evolve under compression.

\subsection{Structural Evolution with Pressure}

Bi-2212 can be viewed as alternating CuO$_2$ layers and charge-reservoir layers \cite{ref30}, as illustrated in Fig.~1(a). The structural response of Bi-2212 to pressure is distinctive. Unlike many other cuprates, which often undergo significant structural transitions under pressure or doping, Bi-2212 has been reported to retain its orthorhombic perovskite structure with the Amaa space group across a wide pressure range \cite{ref5,ref31,ref32,ref33}. Its unit cell is nearly tetragonal (with $a-b\approx 0.1$~\AA) and can be effectively approximated using the I4/mmm tetragonal symmetry. In our calculations, to balance accuracy and computational cost, we employed a body-centered tetragonal structure for Bi-2212 with the I4/mmm space group.

To establish the structural response under pressure, we calculate the lattice parameters of Bi-2212 up to 100 GPa, and the results are shown in Fig.~1(c). The lattice parameters decrease monotonically with increasing pressure, and no isostructural phase transition is found, consistent with previous experimental measurements \cite{ref32}.

\subsection{Electronic Structure under Pressure}

Cuprates are prototypical strongly correlated electron systems, and the states near the Fermi level contain strongly correlated Cu $3d$ electrons. It is therefore essential to include correlation effects in the electronic-structure calculations. In the present first-principles treatment, the onsite Coulomb interaction of Cu $3d$ electrons is included within the DFT+$U$ method with $U_{\mathrm{eff}}=5$ eV, following previous theoretical work \cite{ref34}.

\begin{figure*}[t]
\centering
\includegraphics[width=\textwidth]{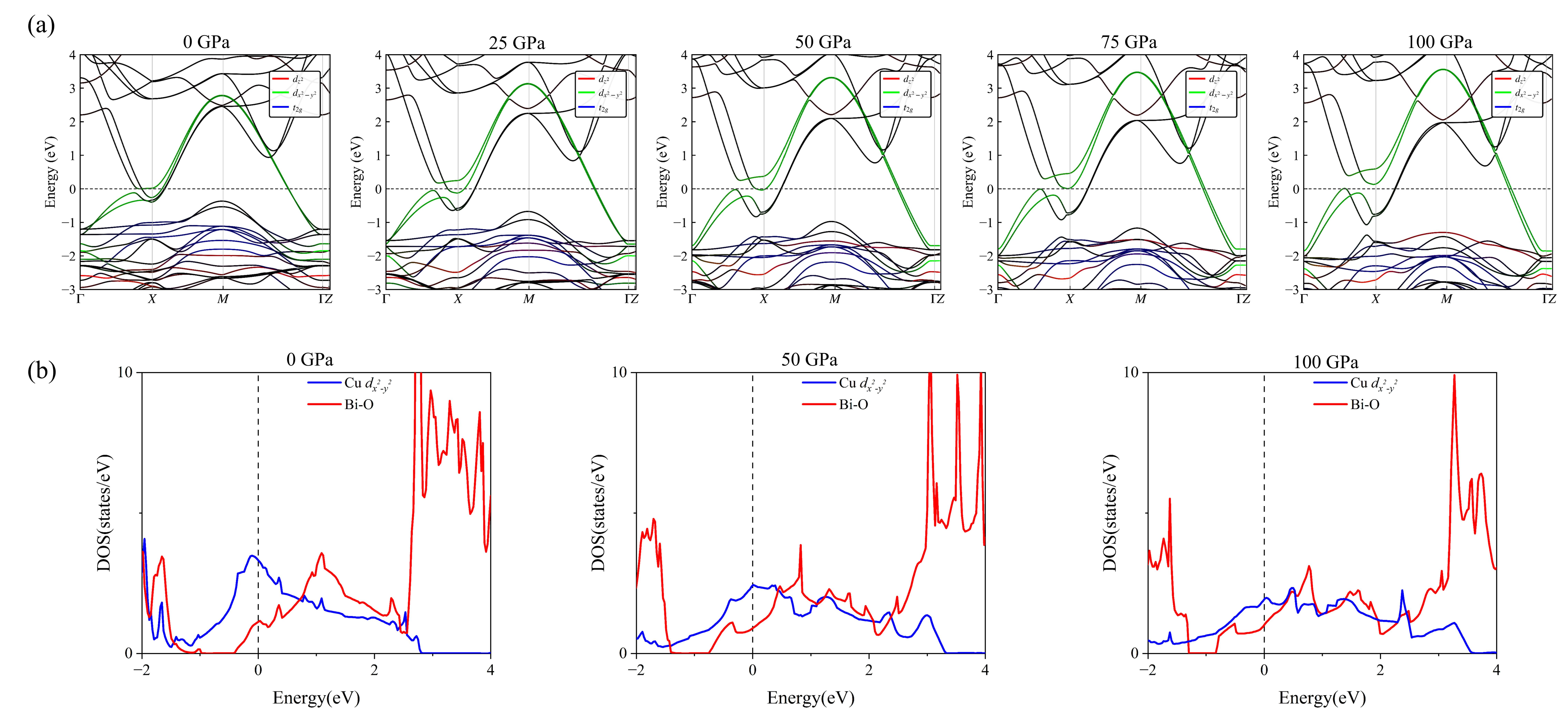}
\caption{DFT+$U$ electronic structure of Bi-2212 for the nominal $\delta=0.5$ case at selected pressures. (a) Band structures at $P=0$, 25, 50, 75, and 100 GPa. (b) Corresponding densities of states at $P=0$, 50, and 100 GPa. The onsite interaction is $U_{\mathrm{eff}}=5$ eV.}
\label{fig:figure2}
\end{figure*}

Using the optimized crystal structures, we compute the electronic structure of Bi-2212 at various pressures. The resulting band structures and densities of states (DOS) for the nominal $\delta=0.5$ case are shown in Fig.~2. The states near the Fermi level contain both Bi--O-derived contributions around X and Cu--O-derived contributions around M. The presence of Bi--O states at the Fermi level gives rise to a self-doping effect \cite{ref35} in Bi-2212, indicating that not all carriers introduced by hole doping reside in the superconducting CuO$_2$ planes. The carrier density relevant to superconductivity must therefore be determined self-consistently from the electronic structure. These features are consistent with ARPES measurements of the normal state in Bi-2212 \cite{ref36,ref37,ref38,ref39}.

As pressure increases, both the Bi--O and Cu--O hybridized bands broaden, consistent with the usual pressure-induced increase of bandwidth, as illustrated in Fig.~2(a). The DOS in Fig.~2(b) further allows us to quantify how carriers are partitioned between the Cu--O-derived band and the Bi--O-derived band. This carrier redistribution constitutes a key control parameter for the superconducting transition temperature $T_{c}$. For simplicity, we neglect additional pressure-induced interband hybridization effects in the following analysis.

Figure~3 displays the pressure evolution of the Cu--O-derived Fermi surface and of the effective CuO$_2$-plane hole concentration $\delta_x$ for the nominal $\delta=0.5$ case. As shown in Figs.~3(a--e), increasing pressure redistributes carriers between the Bi--O and Cu--O bands: holes are progressively transferred from the Bi--O band into the Cu--O band, leading to a systematic increase in $\delta_x$ and a concomitant reconstruction of the Cu--O-derived Fermi-surface topology. For completeness, Fig.~3(f) presents the pressure dependence of $\delta_x$ for several nominal doping levels $\delta$. We use $\delta=0.5$ as a representative nominally optimal sample. At ambient pressure, the extracted effective CuO$_2$-plane hole concentration is $\delta_x=0.18$ per Cu site, consistent with previous experimental and theoretical estimates for Bi-2212 \cite{ref16,ref17,ref18,ref40}. Moreover, within the present rigid-band treatment, $\delta=0.5$ corresponds to the experimentally optimal oxygen content $x=0.25$ \cite{ref35}, since one excess oxygen atom approximately removes two electrons from the system.

Furthermore, Fig.~3(f) shows that $\delta_x$ increases with pressure across all nominal dopings. This trend reflects a broader property of cuprates under compression, as also reported in Hg-based cuprates \cite{ref41}. Notably, the effective CuO$_2$-plane hole concentration does not scale linearly with the nominal doping $\delta$, underscoring the role of self-doping. This reinforces the need to determine the carrier density self-consistently from first principles. In particular, $\delta_x$ serves as a key quantitative input for predicting the pressure evolution of $T_{c}$ in our model.

\begin{figure*}[t]
\centering
\includegraphics[width=\textwidth]{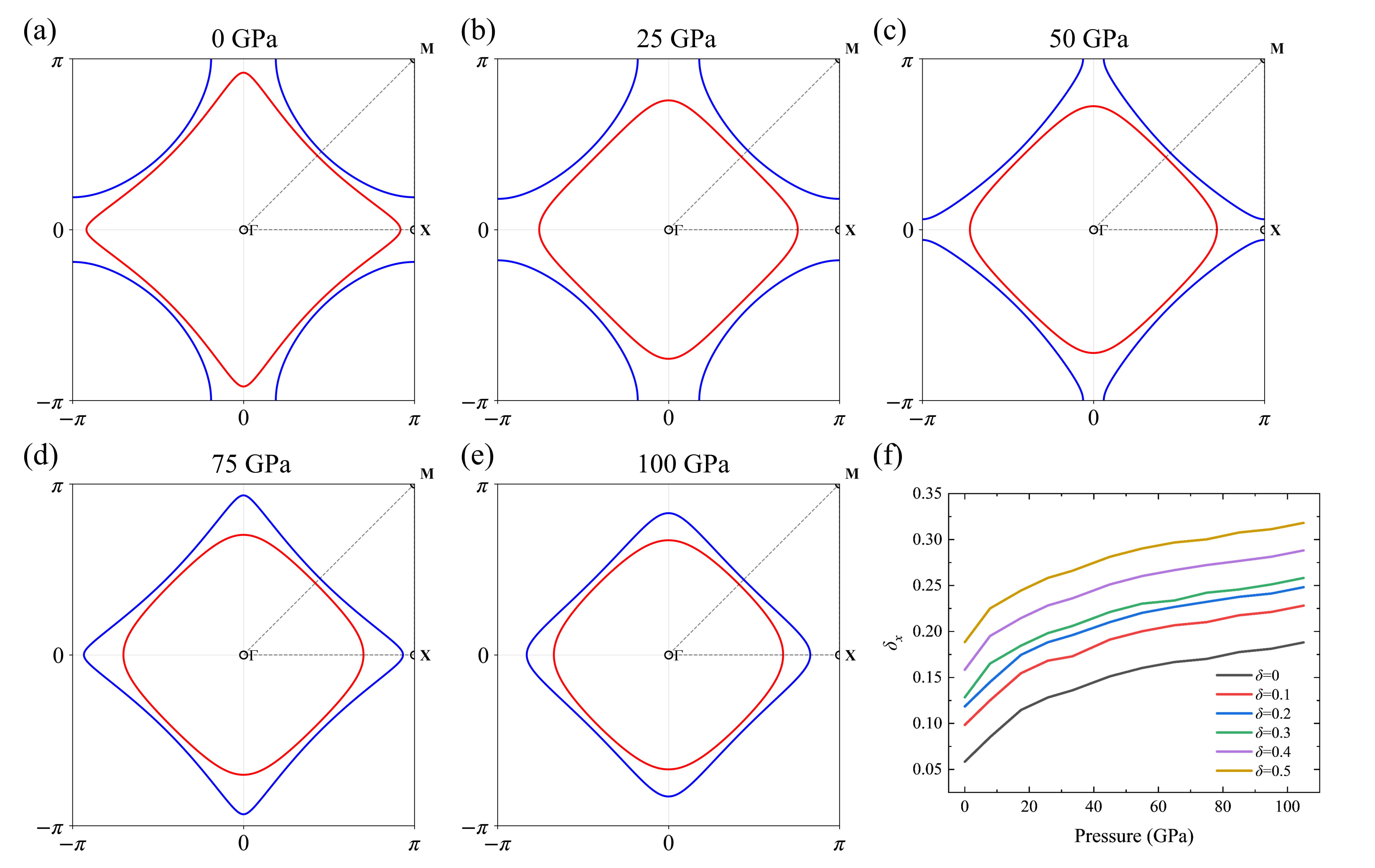}
\caption{(a--e) Cu--O-derived Fermi surfaces of Bi-2212 from DFT+$U$ for the nominal $\delta=0.5$ case at $P=0$, 25, 50, 75, and 100 GPa. The changing enclosed area reflects the pressure-induced variation of the effective CuO$_2$-plane hole concentration. (f) Pressure dependence of the effective CuO$_2$-plane hole concentration $\delta_x$ for several nominal dopings $\delta$.}
\label{fig:figure3}
\end{figure*}

\subsection{The Low-Energy Effective Model}

To investigate the superconducting properties of Bi-2212, we construct a low-energy effective model based on its electronic structure. Using Wannier-function fitting, we extract the effective intralayer hopping parameters $t$ and $t'$. Building on this foundation, we incorporate the nearest-neighbor intralayer spin exchange interaction $J$, thereby obtaining the low-energy effective $t-J$ model given in Eq. (3).

\begin{figure*}[t]
\centering
\includegraphics[width=\textwidth]{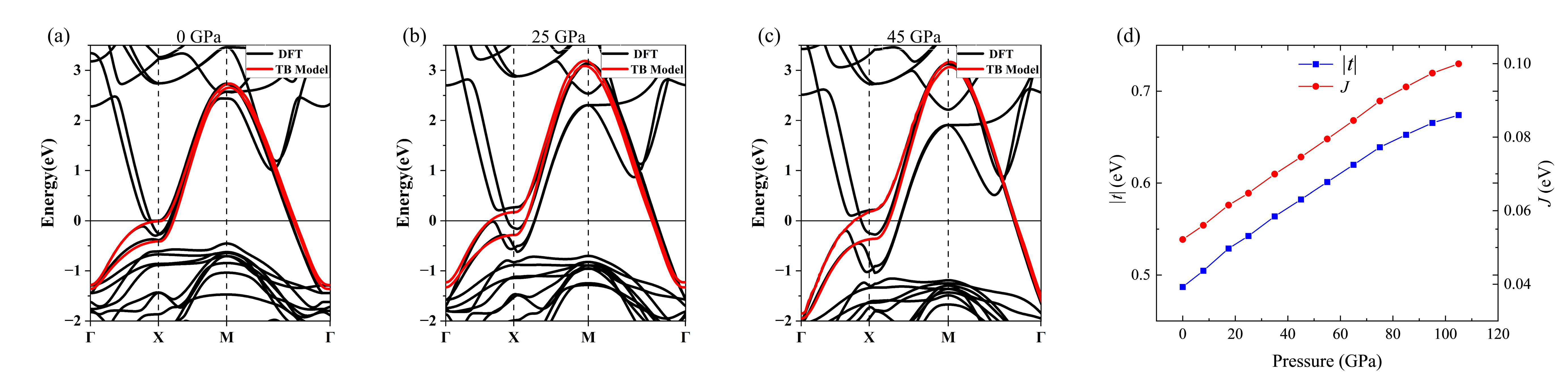}
\caption{(a--c) Comparison between the DFT bands of Bi-2212 and the corresponding bilayer tight-binding/Wannier fits for the nominal $\delta=0.5$ case at $P=0$, 25, and 45 GPa. The fitted Cu $d_{x^{2}-y^{2}}$ bands are shown as red lines. (d) Pressure dependence of the magnitude $|t|$ of the nearest-neighbor Wannier hopping and the superexchange parameter $J$.}
\label{fig:figure4}
\end{figure*}

\begin{table*}[t]
\caption{Raw tight-binding parameters of nominal $\delta=0.5$ Bi-2212 extracted from Wannier fits at selected pressures. $E_{0}$ denotes the onsite energy of the Cu $d_{x^{2}-y^{2}}$ orbital; $t$ and $t'$ denote the intralayer nearest-neighbor and next-nearest-neighbor hopping parameters, respectively, with the sign convention of the Wannier fit retained; and $t_{\perp}$ denotes the interlayer nearest-neighbor hopping parameter.}
\label{tab:tightbinding}
\begin{ruledtabular}
\begin{tabular}{lrrrr}
Pressure & $E_{0}$ & $t$ & $t'$ & $t_{\perp}$ \\
0 GPa & 0.373 & -0.479 & 0.119 & -0.058 \\
25 GPa & 0.519 & -0.542 & 0.109 & -0.089 \\
45 GPa & 0.565 & -0.582 & 0.102 & -0.108 \\
65 GPa & 0.615 & -0.619 & 0.096 & -0.132 \\
85 GPa & 0.639 & -0.652 & 0.090 & -0.151 \\
105 GPa & 0.652 & -0.674 & 0.085 & -0.180 \\
\end{tabular}
\end{ruledtabular}
\end{table*}

The DFT band structures in Fig.~4(a--c) are compared with the corresponding Wannier fits after isolating the Cu--O-derived low-energy manifold. The hybridization with the Bi--O bands is omitted in this comparison so that the Cu $d_{x^{2}-y^{2}}$ bilayer sector can be mapped onto an effective single-orbital model more transparently. The Wannier-fitted dispersions reproduce the DFT Cu--O bands very well, and the extracted raw tight-binding parameters are listed in Table~\ref{tab:tightbinding}. The sign convention of the Wannier fit is retained in the table; accordingly, panel (d) shows the magnitude $|t|$ rather than $t$ itself. In the low-pressure regime, these parameters are consistent with previous first-principles studies \cite{ref42}.

Following previous theoretical studies on cuprates \cite{ref21,ref43}, the nearest-neighbor exchange interaction $J$ is taken to scale as $4t^{2}/U$, with $U$ treated as pressure independent. We then calibrate the overall scale of $J$ by requiring that the ambient-pressure maximum of the calculated $T_{c}$--$\delta_x$ dome be $80$ K. The resulting pressure dependence of $J$ should therefore be regarded as a semi-empirical input anchored to experiment rather than as a fully parameter-free first-principles quantity. As shown in Fig.~4(d), $J$ increases with pressure, indicating that compression enhances the pairing interaction and can substantially raise the intrinsic pairing scale of Bi-2212.

\subsection{Evolution of the $T_{c}$}

Having established the pressure dependence of the model parameters and the CuO$_2$-plane hole concentration $\delta_x(P)$, we now analyze $T_{c}$ as a joint function of pressure and initial doping. Figs.~5(a), 5(b), and 6 show that Bi-2212 under pressure cannot be characterized by a single, sample-independent $T_{c}(P)$ curve. Instead, the superconducting response is governed by how pressure-driven self-doping moves a given sample across the common $T_{c}$--$\delta_x$ dome.

As shown in Fig.~5(a), the $T_{c}$--$\delta_x$ relation preserves the familiar dome-shaped form at all representative pressures, with the optimal CuO$_2$-plane hole concentration remaining near $\delta_x\approx 0.18$, consistent with previous experimental data \cite{ref16,ref17,ref18,ref40} and with our first-principles estimates. At the same time, the height of this dome increases with pressure. Combined with the pressure dependence of the effective hopping and superexchange parameters, this indicates that compression enhances the intrinsic pairing scale and raises the maximum attainable $T_{c}$. Pressure therefore affects superconductivity through two intertwined channels: it lifts the $T_{c}$ dome vertically by strengthening the pairing interaction, and it shifts the sample horizontally along the dome by increasing the CuO$_2$-plane hole concentration through self-doping.

Once these two effects are distinguished, the strong initial-doping sensitivity of $T_{c}(P)$ follows naturally. For a sample starting on the underdoped side, pressure transfers holes from the Bi--O reservoir to the CuO$_2$ planes and drives the system toward optimal doping, while simultaneously increasing the pairing strength. In this case, the two effects cooperate and $T_{c}$ increases with pressure. By contrast, for a sample starting near ambient-pressure optimal doping, the same self-doping mechanism quickly shifts the system to the overdoped side of the dome. The horizontal displacement away from the optimal region can then offset, and eventually outweigh, the pressure-induced enhancement of the pairing scale. As a result, $T_{c}$ may exhibit only a weak enhancement, a plateau, or an overall suppression under pressure.

\begin{figure*}[t]
\centering
\includegraphics[width=\textwidth]{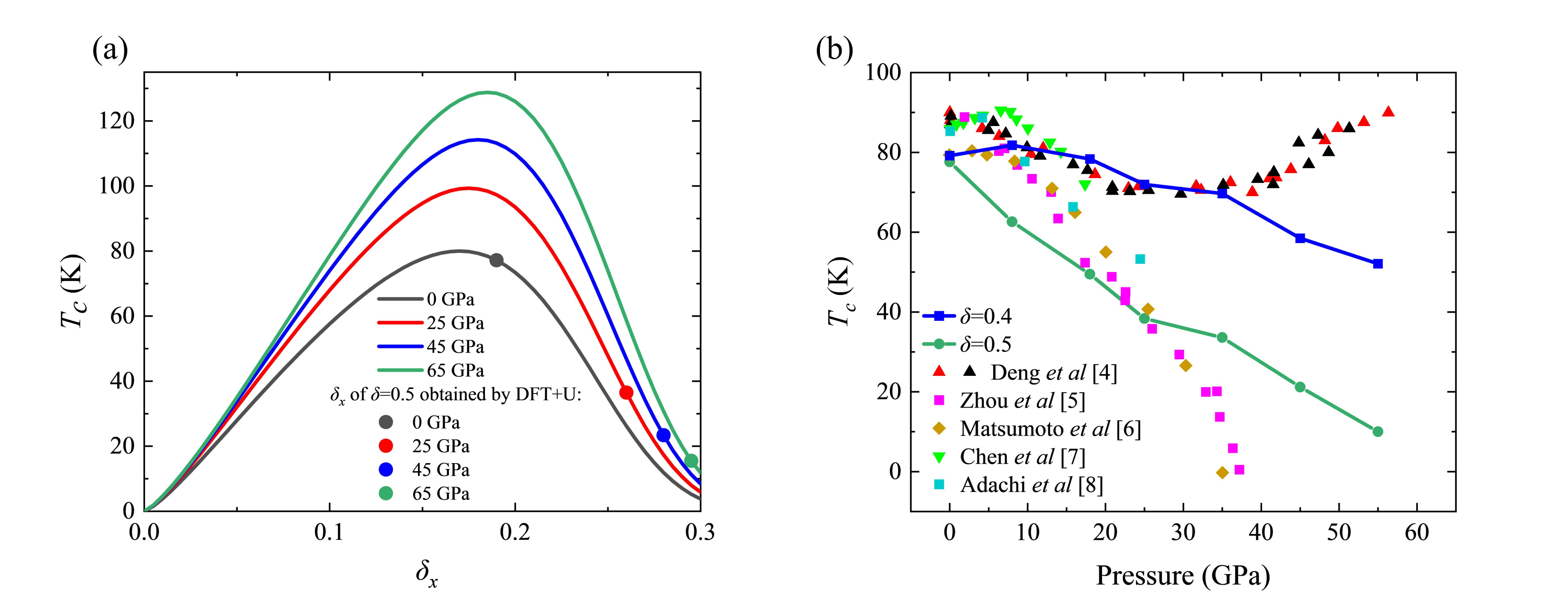}
\caption{(a) Calculated $T_{c}$--$\delta_x$ curves of Bi-2212 at selected pressures, obtained from the SBMF+BKT scheme. The colored dots mark the DFT-derived CuO$_2$-plane hole concentration $\delta_x(P)$ for the nominal $\delta=0.5$ sample. (b) Calculated $T_{c}(P)$ curves for two nominally near-optimal samples, $\delta=0.4$ and $\delta=0.5$, compared with representative experimental data for near-optimal Bi-2212 under pressure \cite{ref4,ref5,ref6,ref7,ref8}.}
\label{fig:figure5}
\end{figure*}

Fig. 5(b) makes the strong initial-doping sensitivity of the pressure response more explicit by comparing two nominally near-optimal trajectories, $\delta = 0.4$ and $\delta = 0.5$. Although the two cases have similar $T_{c}$ values at ambient pressure, their evolutions under compression are markedly different. The $\delta = 0.5$ curve is rapidly suppressed with increasing pressure because pressure-induced self-doping quickly drives the CuO$_2$-plane hole concentration $\delta_x$ to the overdoped side of the $T_{c}$-$\delta_x$ dome, so that the horizontal shift away from the optimum outweighs the pressure-enhanced pairing scale. By contrast, the $\delta = 0.4$ curve remains in the high-$T_{c}$ region over a much broader pressure interval and shows only a moderate reduction, indicating that a slightly lower initial doping can better compensate the carrier transfer caused by pressure.

The comparison with experiment in Fig. 5(b) further shows that even samples all described as near optimally doped do not follow a unique $T_{c}(P)$ curve. The more weakly suppressed $\delta = 0.4$ trajectory is qualitatively closer to the high-$T_{c}$ or weak-suppression class of measurements, represented most clearly by Deng \textit{et al.} \cite{ref4} and partly by Chen \textit{et al.} \cite{ref7}, whereas the more rapidly suppressed $\delta = 0.5$ trajectory is more comparable to the strong-suppression trends reported by Zhou \textit{et al.} \cite{ref5} and Matsumoto \textit{et al.} \cite{ref6}, with the data of Adachi \textit{et al.} \cite{ref8} lying between these two tendencies in the low- and intermediate-pressure range. Although the present calculation does not reproduce all experimental details, especially the high-pressure re-entrant increase reported in Ref.~\cite{ref4}, it clearly demonstrates that small differences in the initial doping level, even within the near-optimal regime, can produce qualitatively different pressure evolutions of $T_{c}$.

Fig. 6 makes the doping selectivity even more transparent. Different initial doping levels produce qualitatively different $T_{c}(P)$ curves: underdoped samples show strong enhancement under compression, moderately doped samples develop a broad high-$T_{c}$ region, whereas samples starting closer to optimal or overdoped conditions are progressively suppressed. More importantly, samples with very similar ambient-pressure $T_{c}$ can still evolve in dramatically different ways once pressure is applied. For example, the $\delta = 0.4$ and $\delta = 0.5$ cases start from comparable $T_{c}$ values at ambient pressure, yet they diverge strongly under compression, with the $\delta = 0.5$ sample being suppressed much more rapidly. Therefore, ambient-pressure $T_{c}$ alone is not a sufficient descriptor of the high-pressure response; what matters more fundamentally is the initial position of the sample on the common $T_{c}$--$\delta_x$ dome.

An important implication of these results is that the best starting point for pressure-enhanced superconductivity is not the ambient-pressure optimal sample, but an initially underdoped one. In particular, a slightly underdoped sample offers the most practical compromise: it already possesses a substantial $T_{c}$ at ambient pressure, yet still retains enough room for pressure-induced self-doping to drive the CuO$_2$ planes toward the optimal doping region. More strongly underdoped samples can in principle reach even higher $T_{c}$ values at sufficiently high pressure, but they require a much larger pressure tuning range. From the viewpoint of high-pressure experiments, slightly underdoped Bi-2212 should therefore be a more favorable starting point than an ambient-pressure optimal sample.

\begin{figure}[t]
\centering
\includegraphics[width=0.95\columnwidth]{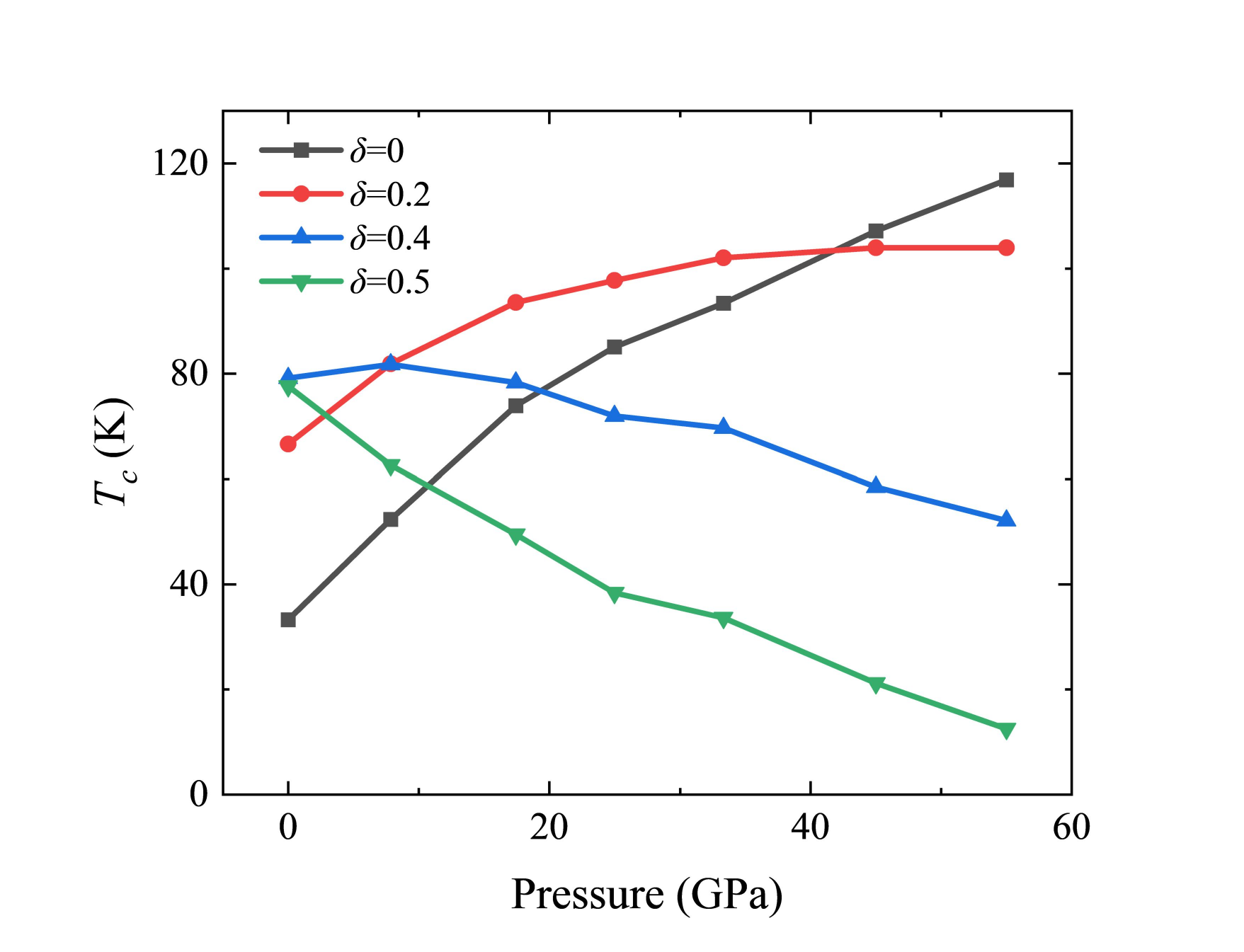}
\caption{Calculated $T_{c}(P)$ curves of Bi-2212 for several initial nominal dopings, $\delta=0$, 0.2, 0.4, and 0.5, defined at ambient pressure. Although some samples have similar ambient-pressure $T_{c}$ values, their pressure evolutions differ strongly, showing that the high-pressure response is governed by the initial position of the sample on the common $T_{c}$--$\delta_x$ dome.}
\label{fig:figure6}
\end{figure}

\section{Discussion and Conclusion}

The central conclusion of the present work is that hydrostatic pressure does not generate a unique, sample-independent $T_{c}(P)$ curve in Bi-2212. Instead, pressure and doping are intrinsically entangled through a pressure-induced self-doping effect. Our DFT calculations show that compression systematically transfers carriers from the Bi--O reservoir to the CuO$_2$ superconducting planes, thereby increasing the effective hole concentration relevant to superconductivity. When this carrier redistribution is combined with the dome-shaped $T_{c}$--$\delta_x$ relation obtained from the SBMF+BKT analysis, a unified physical picture emerges: pressure can either enhance or suppress $T_{c}$, depending on the initial doping state of the sample.

This picture provides a natural way to organize a large part of the apparent disagreement among high-pressure experiments on Bi-2212. Opposite signs of $dT_{c}/dP$ do not necessarily imply different microscopic mechanisms. To a large extent, they can arise because different samples start from different positions on the same superconducting dome. An underdoped sample can exhibit a positive pressure coefficient because pressure both strengthens the pairing scale and moves the system toward optimal doping. By contrast, a sample starting near ambient-pressure optimal doping can display only limited enhancement or even an overall suppression because self-doping immediately drives it toward the overdoped side. In this sense, the apparently conflicting pressure dependences reported in Refs.~\cite{ref4,ref5,ref6,ref7,ref8} should not simply be viewed as mutually contradictory observations, but rather as different trajectories associated with different initial carrier concentrations.

A second key message is that the relevant control parameter is not merely the nominal chemical doping, but the effective hole density in the CuO$_2$ superconducting layers under pressure. Because Bi-2212 contains a charge-reservoir block, nominally introduced carriers are not confined to the CuO$_2$ planes, and pressure further redistributes this charge. The distinction between nominal doping and superconducting-layer doping is therefore essential for understanding the pressure response. This also explains why samples with similar ambient-pressure $T_{c}$ can behave very differently once compressed. In this respect, the pressure-induced self-doping effect identified here is a key microscopic ingredient that has to be taken into account when interpreting Bi-2212 under pressure.

Our results also suggest a practical strategy for future high-pressure experiments. If the goal is to maximize the superconducting performance of Bi-2212 under pressure, it is not sufficient to begin with samples that are optimal at ambient pressure. A slightly underdoped starting point is more advantageous, because it leaves room for pressure-induced self-doping to move the system toward the top of the dome while still benefiting from the pressure-enhanced pairing interaction. Thus, the present work not only offers an interpretation of existing experimental discrepancies, but also provides a concrete guide for sample selection in future pressure studies.

At the same time, initial-doping sensitivity is likely not the whole story in the most extreme pressure regime. More anomalous reports, such as a possible second superconducting dome \cite{ref4} or insulating-like behavior at high pressure \cite{ref5}, may involve additional ingredients beyond the hydrostatic self-doping mechanism identified here. Possible factors include non-hydrostatic stress, pressure-induced oxygen redistribution, pressure-enhanced disorder, or other electronic reconstructions. The present results should therefore be regarded as a unified explanation for the dominant pressure trends in Bi-2212, while leaving room for additional physics in the most extreme high-pressure situations.

In summary, by combining first-principles calculations with an effective low-energy model and an SBMF+BKT treatment of superconductivity, we have established a microscopic framework for understanding the evolution of $T_{c}$ in Bi-2212 under pressure. Pressure has a dual effect: it enhances the effective pairing scale, but it also drives self-doping that increases the hole concentration in the CuO$_2$ planes. Because $T_{c}$ remains dome-shaped as a function of the CuO$_2$-plane carrier density, the competition between these two effects makes the pressure response extraordinarily sensitive to the initial doping state. This explains why Bi-2212 can exhibit either a pressure-induced enhancement or a pressure-induced suppression of $T_{c}$, and why slightly underdoped samples are the most promising route for achieving improved superconducting performance under pressure.

\begin{acknowledgments}
This work is supported by the National Natural Science Foundation of China (NSFC) under Grant No. 11974354. Numerical calculations were partly performed in the Center for Computational Science of CASHIPS, the ScGrid of Supercomputing Center, Computer Network Information Center of Chinese Academy of Sciences, and the Hefei Advanced Computing Center.
\end{acknowledgments}

\end{document}